\documentclass[%
 reprint,
superscriptaddress,
 amsmath,amssymb,
 aps,
prl,
floatfix,
]{revtex4-2}

\usepackage{graphicx}
\usepackage{dcolumn}
\usepackage{bm}
\usepackage{cancel}%
\usepackage{xcolor}
\usepackage[normalem]{ulem}
\usepackage{subfiles}

\usepackage{graphics}
\usepackage{enumerate}
\usepackage{epsfig}
\usepackage{amsmath}
\usepackage{amssymb}
\usepackage{physics}
\usepackage{comment}
\usepackage[caption=false]{subfig}
\usepackage{hyperref}
\usepackage{float}
\usepackage{xcolor}
\usepackage{afterpage}
\usepackage[normalem]{ulem}
\usepackage[counter,user,hyperref]{zref}
\usepackage{cleveref}
\usepackage{lipsum}

\def\VR{\kern-\arraycolsep\strut\vrule &\kern-\arraycolsep}
\def\vr{\kern-\arraycolsep & \kern-\arraycolsep}

\newcommand{\mlm}[1]{{\leavevmode\color{blue}[#1]}}

\begin{document}

\title{Free-then-freeze: transient learning degrees of freedom for introducing function in materials}

\author{Varda F. Hagh}
\affiliation{James Franck Institute, University of Chicago, Chicago, IL 60637, USA}

\author{Sidney R. Nagel}
\affiliation{James Franck Institute, University of Chicago, Chicago, IL 60637, USA}

\author{Andrea J. Liu}
\affiliation{Department of Physics, University of Pennsylvania, Philadelphia, PA 19104, USA}

\author{M. Lisa Manning}
\affiliation{Department of Physics and BioInspired Institute, Syracuse University, Syracuse, NY 13244, USA}

\author{Eric I. Corwin}
\affiliation{Department of Physics and Materials Science Institute, University of Oregon, Eugene, OR 97403, USA}

\date{\today}

\begin{abstract}
The introduction of transient learning degrees of freedom into a system can lead to novel material design and training protocols that guide a system into a desired metastable state. In this approach, some degrees of freedom, which were not initially included in the system dynamics, are first introduced and subsequently removed from the energy minimization process once the desired state is reached.
Using this conceptual framework, we create stable jammed packings that exist in exceptionally deep energy minima marked by the absence of low-frequency quasilocalized modes; this added stability persists in the thermodynamic limit. 
The inclusion of particle radii as transient degrees of freedom leads to deeper and much more stable minima than does the inclusion of particle stiffnesses. This is because particle radii couple to the jamming transition whereas stiffnesses do not. Thus different choices for the added degrees of freedom can lead to very different training outcomes.

\end{abstract}

\maketitle
\subsection*{Introduction}
Systems with disorder often have a complex and rugged energy landscape whose minima are determined by the system’s degrees of freedom and constraints. Finding low-lying states in such a system is a daunting challenge that lies at the heart of many constraint-satisfaction statistical-physics problems, ranging from machine learning to population ecology~\cite{krzakala2007landscape,mezard2009constraint,altieri2019constraint,mehta2019constrained}.  More difficult is finding rare minima in such systems that have specific desired properties or functions such as enhanced stability against external perturbations. 
Finding a framework for designing desired function into a complex system in a flexible and reliable manner is an important goal that could lead to a paradigm shift in material processing. 

There is an unappreciated but common feature of many of the protocols that have recently been proposed to create ground states with special properties. In each case, new ``learning" degrees of freedom, not contained in the original description, are added to the system. That is, some constraints are relaxed and allowed to vary according to a set of dynamical rules. These degrees of freedom are then manipulated to produce the desired behavior before being removed. We term these "free-then-freeze" optimization protocols.

To make this explicit, we give a few examples. \textit{Tuning-by-pruning} protocol for dilution of bonds in an elastic network~\cite{goodrich2015principle,hagh2018disordered,hagh2019broader, hexner2018linking, hexner2018role}: here the new degree of freedom is the possibility of removing a bond. \textit{Directed-aging} protocol for evolving  function by aging a material under imposed strains~\cite{pashine2019directed,  hexner2020effect,hexner2020periodic}: the new degree of freedom in this case, is the evolution of the stiffness or length of each bond. \textit{Swap Monte-Carlo protocol} for allowing long-ranged exchange of particles with different sizes~\cite{ninarello2017models,ikeda2017mean, brito2018theory, berthier2019efficient, parmar2020stable}: the new degrees of freedom are the particle swaps. More generally, neural networks are tuned by varying node or edge properties to learn tasks. 

Introducing new learning degrees of freedom alters the energy landscape and, as some of these protocols have demonstrated, allows states with rare and desired properties to be accessed. Moreover, these degrees of freedom are transient: they are accessible during the system evolution but are subsequently frozen.  They can be removed either explicitly (by freezing them after the training has been completed) or implicitly (by noting that a separation of time scales can appear naturally during evolution so that some relaxations are no longer possible).  

While free-then-freeze optimization protocols have been studied in a wide range of networks or thermal systems, here we focus on athermal particulate materials, which pose an additional challenge to introducing function. In such systems, states with desired properties must correspond to local minima. Such minima are often explored via localized rearrangements that alter the topology of the connection between nodes. This can lead to long-range elastic stresses and reorganization within the material via avalanches, making it difficult to control final properties.

Jammed particle packings are a paradigmatic category of athermal particulate systems. The complex landscape of an $N$-particle packing in $d$-dimensions is the $Nd-1$ potential-energy surface in the $Nd$ space of the translational degrees of freedom of particles associated with the positions of each of the particles. Typically, in the thermodynamic limit a jammed system is only \textit{marginally stable} to perturbations~\cite{Connelly1996, charbonneau2012universal,goodrich2013stability, goodrich2014jamming, degiuli2014effects, Connelly2015, muller2015marginal, xu2017instabilities}; that is, an infinitesimal perturbation suffices to rearrange the packing, pushing the system into a new local energy minimum. Because material stability is a key property governing the response of a material under shear and deformation, such as the transition from ductile to brittle behavior~\cite{yeh2020glass}, producing jammed packings with high mechanical stability has been an active area of interest~\cite{ninarello2017models,ikeda2017mean, brito2018theory, berthier2019efficient, parmar2020stable, brito2018universality, kapteijns2019fast,ikeda2019mean, yanagishima2020towards, ikeda2020infinitesimal}. 

Some of these works~\cite{brito2018universality,brito2018theory} have started to develop organizing principles to explain how free-then-freeze optimization, specifically focused on changing particle radii, can drive high stability in jammed packings. However, a broader understanding of which variables are important to constrain, and why, has remained elusive.
Here we introduce a new set of global free-then-freeze protocols and apply them to jammed packings. We contrast how involving two different sets of learning degrees of freedom affect material stability. We find that not all learning degrees of freedom are equally useful for evolving a desired function, and show that this can be explained by an analysis of how the constraints enter into the dynamical equations. 

Manipulating learning degrees of freedom in this transient way accesses local minima that not only have low energy but also have high energy barriers preventing escape. By understanding which variables alter stability the most, we can better understand the factors that control the effectiveness of specific preparation protocols, potentially opening up new avenues for the rational design and manipulation of materials.


\subsection*{Protocol Details}

\begin{figure}[h!]
\centering
\includegraphics[trim= 0cm 0cm 0cm 0cm, width=8cm]{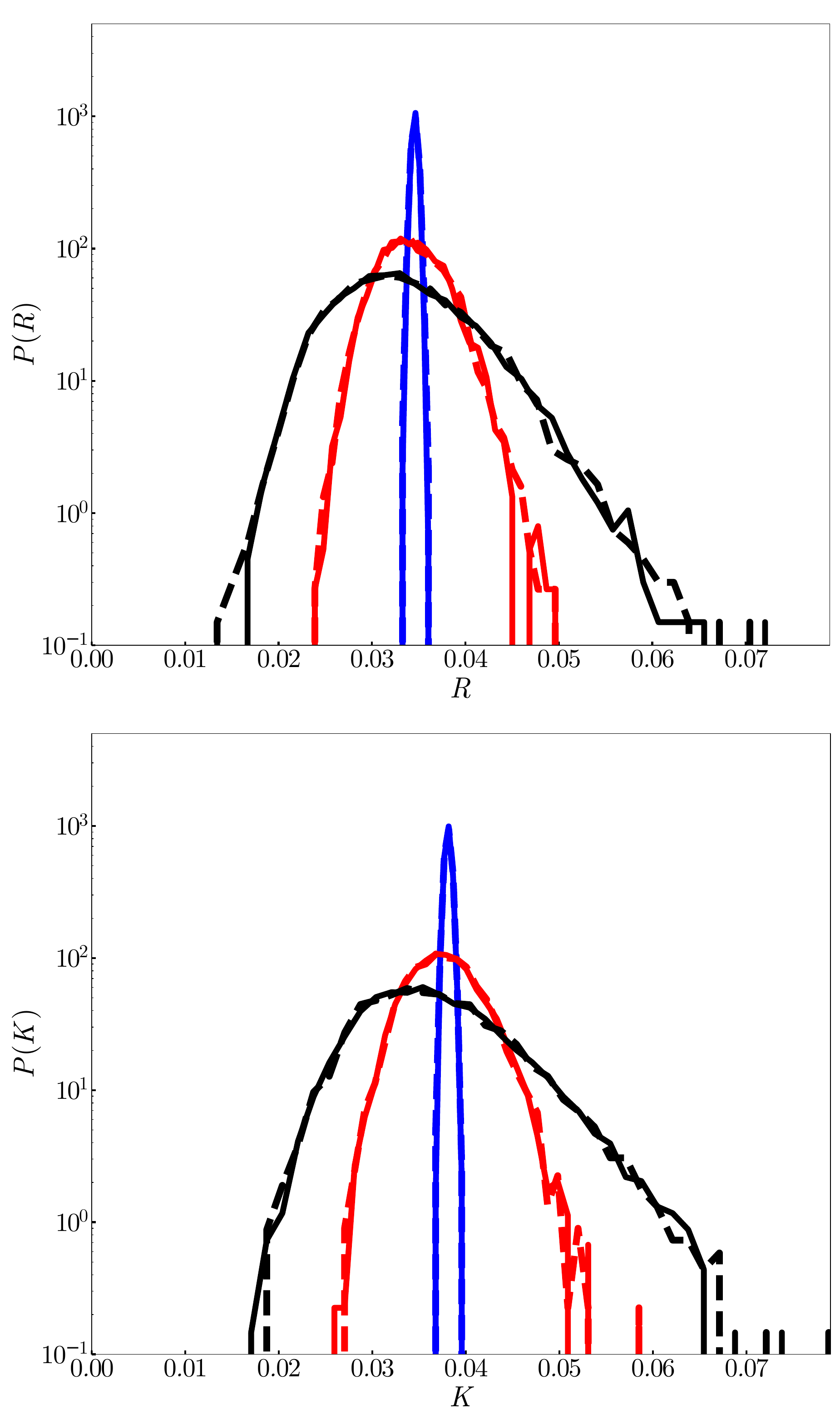}%
\caption{Distributions of radii (top) and stiffnesses (bottom) before (solid) and after (dashed) minimization in $3D$ packings. The different distributions have different fractional widths about their average value: $1 \%$ (blue), $10 \%$ (red), and $20 \%$ (black). Each packing has $N=4096$ particles and is prepared at a pressure of $p = 10^{-2}$. 
}
\label{fig:distributions}
\end{figure}   

We study packings of spheres in three dimensions, $3D$.  Particles, labeled by $i$ and $j$ interact via finite-range, repulsive harmonic potentials:
\begin{equation}
    U = \frac{1}{2} \ \sum_{i,j} \ \epsilon_{ij} \ \left( 1 - \frac{|\mathbf{X}_i - \mathbf{X}_j |}{R_i + R_j} \right)^2 \Theta \left( 1 - \frac{|\mathbf{X}_i - \mathbf{X}_j |}{R_i + R_j} \right),
    \label{eq:energy}
\end{equation}
where $\mathbf{X}_i$ and $R_i$ are the position and radius of particle $i$, and $\epsilon_{ij}= \frac{K_i K_j}{K_i + K_j}$ is the interaction stiffness between particles $i$ and $j$. As is standard, the individual stiffnesses of particles $K_i$ and $K_j$ are added like springs in series to produce the effective interaction between them. The Heaviside step function allows only positive overlaps between particles to contribute.

We generate jammed packings by randomly distributing $N$ particles at a given volume fraction, $\phi$, in a cubic box of linear size $L \equiv 1$ and then minimizing the energy with respect to the degrees of freedom~\cite{durian_foam_1995-1, o2003jamming}.
In a conventional packing, the only degrees of freedom are the particle positions. Here we consider two additional sets of degrees of freedom, namely the particle radii, $ \{R_{i} \}$, and stiffnesses, $\{ K_{i} \}$. We minimize the energy with respect to the degrees of freedom using pyCUDAPacking~\cite{charbonneau2012universal,morse2014geometric}, incorporating a quad-precision GPU implementation of the FIRE algorithm~\cite{bitzek2006structural}. 

With either $R_{i}$ or $K_{i}$ as learning degrees of freedom, there are trivial zero-energy minima to which the system can escape. For example, a sufficient number of the radii can shrink in order to remove all overlaps, or the stiffness cost of overlaps can adjust to zero. To avoid such trivial states, we constrain some of the moments, $m$, of the distributions of the new variables so that $\Phi_{\chi,m} \equiv \sum_i \chi_i^m$  are fixed, where $\chi_{i}$ can be either $R_{i}$ or $K_{i}$. For $\chi_{i}=R_{i}$, fixing the packing fraction $\phi \propto \Phi_{R,3} = \sum_i R_i^3$ is not sufficient to prevent some radii from shrinking to zero. By choosing constraints with both negative and positive powers of $\chi_{i}$, namely $m = \{-3, -2, -1, 1, 2, 3, 6 \}$, we are able to minimize all of the systems presented in this paper without appreciably altering the radius or stiffness distributions (Fig.~\ref{fig:distributions}). Note that the set of constraints chosen is not special and one can produce similar results with other $\{m\}$ similarly including both positive and negative powers of $\chi_{i}$. Here, we report data for $3D$ packings with log-normal distributions of initial and final polydispersity of $20 \%$ in $\{R_i\}$ or $\{K_i\}$ (except in Fig.~\ref{fig:distributions} where multiple polydispersities are reported). 

To impose the constraints during minimization, we restrict changes in the new variables to subspaces in which all constraints are satisfied. The subspace of allowed values is the space spanned by vectors $\mathbf{\nabla}_{\chi} (\sum_i \chi_i^m)$.  We identify the unit vectors, $\hat{\mathbf{n}}_{\chi, m}$ that form a complete orthonormal basis for this subspace and project out force components perpendicular to the subspace:
\begin{equation}
   \label{eq:constrainedForce}
   \mathbf{f}_{\chi, \{m\} } = \mathbf{f}_{\chi} -  \sum_m \left( \mathbf{f}_{\chi} \cdot \hat{\mathbf{n}}_{\chi, m} \right) \hat{\mathbf{n}}_{\chi, m .}
\end{equation}
These forces are used in the FIRE minimizer to update radii and stiffnesses; particle positions are updated without any constraints.

\subsection*{Measuring Stability}

\begin{figure*}[t!]%
\centering
\includegraphics[width=18cm]{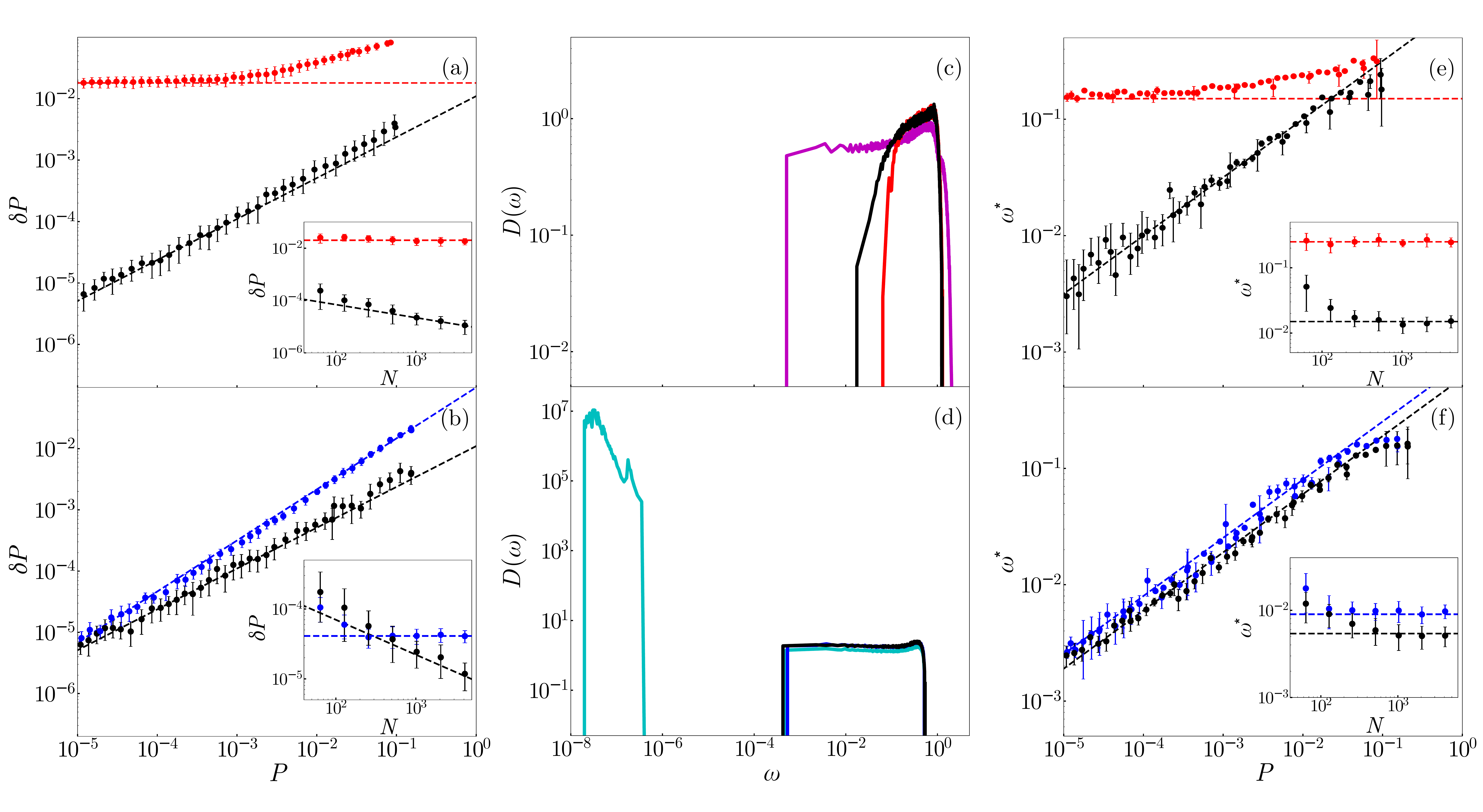}
\caption{\textbf{a, b}) The increase in pressure, $\delta P$, required to make a packing unstable for (a) radius- (red) and (b) stiffness-minimized (blue) packings as well as conventional packings (black) with the same distributions of radii and stiffness. Black dashed lines show a power law $\delta P \propto P^{2/3}$, the red dashed line in (a) is constant (zero slope) and the blue dashed line in (b) shows $\delta P \propto P^{5/6}$. Insets in both panels: $\delta P$ versus system size $N$ at $P=10^{-4}$; black dashed lines indicate $\delta P \propto P^{-1/3}$. 
\textbf{c, d}) Density of states, $\mathcal{D}(\omega)$ versus $\omega$, for (c) radius 
and (d) stiffness-minimized  
packings near jamming transition at $\phi \simeq 0.70$ and $\phi \simeq 0.64$, respectively. The magenta (c) and cyan (d) curves show $\mathcal{D}(\omega)$ when radii and stiffnesses are degrees of freedom, while the red (c) and blue (d) curves show $\mathcal{D}(\omega)$ once radii/ stiffnesses are frozen at their equilibrium values after minimization. The black curves in both plots show the density of states for conventional packings with the same radii and stiffness distributions. 
\textbf{e, f}) $\omega^{*}$ versus pressure $P$ for (e) radius (red) and (f) stiffness-minimized (blue) packings. The black data in both panels represent conventional packings. Black dashed lines in both plots are a power law of $\omega^{*} \propto P^{1/2}$. Insets: ensemble averaged $\omega^{*}$ versus system size, $N$, for packings at pressure $P=10^{-4}$. The lines have zero slope.  Data points in the main plots are averaged over $20$ samples each with $N = 1024$ particles.}
\label{fig:panel}
\end{figure*}

We measure stability in three ways (Fig.~\ref{fig:panel}). Figs.~\ref{fig:panel}a,b present the increase in pressure, $\delta P$, needed to destabilize the system (see the Supplemental Information for details). Fig.~\ref{fig:panel}a shows that radius-minimized packings (red) require a much larger change in pressure to become unstable compared to ordinary packings with the same radius distribution. Note that $\delta P \rightarrow$ \textit{constant} at low pressure for the radius-minimized packings.  This is in contrast to $\delta P \rightarrow 0$ as $P \rightarrow 0$ for ordinary packings, reflecting marginal stability at the jamming transition. 
Fig.~\ref{fig:panel}b shows that the pressure change required to destabilize a stiffness-minimized packing (blue) is also larger than that of a conventionally prepared packing with the same distribution of particle stiffnesses (black) but vanishes as $P \rightarrow 0$. This result shows that the stiffness degrees of freedom become less effective in increasing stability as the pressure decreases. The insets in panels a and b show that for radius- and stiffness-minimized packings, $\delta P$ is independent of system size for large $N$ at $P=10^{-4}$. This suggests that in the thermodynamic limit, one can use this protocol to create jammed packings that are fully stable and therefore energetically rigid~\cite{damavandi2021energetic}, not marginally stable like ordinary jammed packings.

Figs.~\ref{fig:panel}c,d show the density of vibrational modes $\mathcal{D}(\omega)$. Prior work has shown that for jammed packings, the average energy barriers associated with vibrational modes increase monotonically with mode frequency, $\omega$~\cite{xu2010anharmonic}. This suggests that when the low-frequency tail of $\mathcal{D}(\omega)$ shifts to higher values, the system becomes more stable because the energy-barrier heights increase, preventing the system from easily moving from one local minimum to another. 

Fig.~\ref{fig:panel}c shows that when radius degrees of freedom are allowed to vary, the low-$\omega$ modes (magenta data) extend to much lower frequencies than conventionally prepared packings (black data). These low-frequency modes are coupled to positional degrees of freedom and create a plateau in the density of states. Once the radius degrees of freedom are frozen at their equilibrium values (after the minimization has taken place), the shape of $\mathcal{D}(\omega)$ changes dramatically (red data), shifting the low end of the plateau to much higher frequency.

In contrast, when stiffness degrees of freedom are allowed, as shown in Fig.~\ref{fig:panel}d (cyan data), a band of very low-frequency modes appears. These modes are uncoupled to the position variables, which are responsible for the band of modes at higher frequencies that is essentially the same as for ordinary jammed packings (black). Between these two sets of modes there is a pronounced band gap. When the stiffness degrees of freedom are frozen after minimization (blue), however, the low-frequency band disappears, leaving only a density of states that is very similar to that of a conventionally prepared packing (black) with the same stiffness distribution.

Finally, Figs.~\ref{fig:panel}e,f show measurements of the crossover frequency $\omega^*$, marking the low-frequency end of the plateau in the density of vibrational states, for the radius (red) and stiffness-minimized packings (blue) respectively. In the conventionally-prepared jammed packings (black) $\omega^{*} \propto P^{1/2}$~\cite{silbert2005vibrations}. As shown by the red data in Fig.~\ref{fig:panel}e, $\omega^{*}$ approaches a constant as pressure $P$ is lowered for the radius-minimized packings. By contrast, for the stiffness-minimized packings (blue) $\omega^{*} \propto P^{1/2}$ as in conventional packings, albeit with a slightly higher prefactor.  This behavior again points to a clear distinction between the radius and stiffness degrees of freedom.  The radius-minimized packings behave like conventional packings far above the jamming threshold (at much higher pressure) while  stiffness-minimized packings are similar to conventional packings at the same pressure. The insets in Figs.~\ref{fig:panel}e,f show that $\omega^{*} \rightarrow$ \textit{constant} at large $N$ at $P=10^{-4}$, suggesting it is nonzero as $N \rightarrow \infty$.

\subsection*{Different types of learning degrees of freedom}
Fig.~\ref{fig:panel} shows that adding and then freezing either the radius or stiffness degrees of freedom leads to enhanced stability in jammed packings.  However, the stability gained from the radii is qualitatively different from that gained from stiffnesses. This distinction stems from a fundamental difference in how radius and stiffness degrees of freedom affect the onset of rigidity in jammed systems.

\begin{figure}[h!]
\centering
\includegraphics[trim=1cm 1cm 1cm 1cm, width=\linewidth]{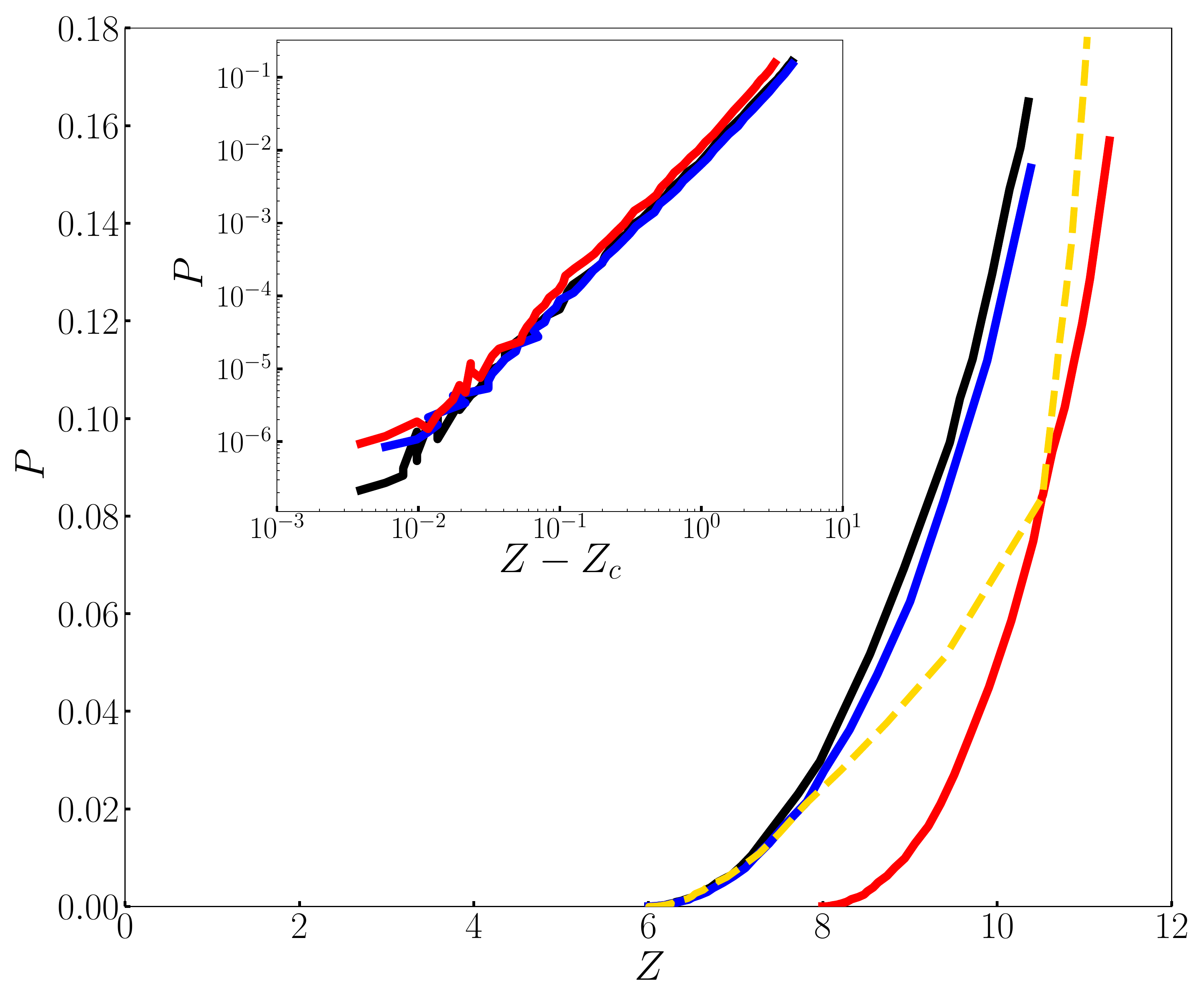}%
\caption{Pressure, $P$, versus coordination, $Z$, for radius- (red) and stiffness- (blue) minimized packings in comparison to conventional (black) packings. The dashed yellow line shows the change in $P$ and $Z$ when a radius-minimized packing is pushed to higher or lower pressures using positions as the only allowed degrees of freedom. The inset shows the pressure against distance from critical point in logarithmic scale. For the red curve, the critical point occurs at $Z_c = 2(d+1)$ while for the blue and black curves $Z_c = 2d$.}
\label{fig:PdeltaZ}
\end{figure}    

This difference is shown in Fig.~\ref{fig:PdeltaZ}, where we plot pressure, $P$, vs.~the number of contacts per particle, $Z$, in radius- and stiffness-minimized packings as well as conventional ones. Introducing radii as degrees of freedom shifts the critical point from $Z_c = 2d$ to a much higher value $Z_c = 2(d+1)$ (ignoring terms of order $1/N$), which lies deep within the stable regime~\cite{degiuli2014effects}. By contrast, stiffness degrees of freedom do not alter the onset of rigidity. Particle stiffnesses become irrelevant to the mechanics of the system as $P \rightarrow 0$ when the particles are just in contact and do not overlap. 

This can be understood from the role that radius and stiffness variables play in the rigidity (compatibility) matrix, $\mathbf{C}$. This is the matrix of coefficients in the linear equations relating changes in the degrees of freedom to changes in the constraints. Here, most of the constraints punish overlaps between particles $h_{ij} =  1 - \frac{|\mathbf{X}_i - \mathbf{X}_j |}{R_i + R_j}$ meaning that $C_{ij} = \partial h_{ij}/ \partial \chi_{i}$, where $\chi_{i}$ can be $X_i$, $R_i$, or $K_i$. Maxwell's constraint count~\cite{Maxwell1864} is then derived using the rank-nullity theorem~\cite{alama2007rank+} on the rigidity matrix $\mathbf{C}$ and its transpose $\mathbf{C}^{T}$. Since the stiffnesses do not affect which particles overlap, they do not appear as independent columns in the rigidity matrix. This means that stiffnesses do not change the rank of the rigidity matrix and therefore cannot change the number of contacts at the transition point, given by the Maxwell's count (see Supplemental Information for further details).

\subsection*{Discussion}

In this paper, we have shown that introducing particle sizes and stiffnesses as transient learning degrees of freedom into the quenching process of jammed packings allows the creation of particularly stable packings with deep minima and high energy barriers against rearrangements. 
However, not every degree of freedom is equally effective in this process; because they do not couple to the jamming transition, the stiffnesses do not affect the stability as profoundly as the particle radii do. 
We also note that while introducing $N$ extra degree of freedom (one per particle), we needed to impose only a few global constraints on the moments of the radius (or stiffness) distributions to ensure that the minimization did not flow to a different fixed point where one particle grew (or became stiffer) at the expense of all the others. For the system sizes considered, $N \le 4096$, we imposed seven constraints. For smaller systems, we needed to impose fewer constraints to obtain similar quantitative results.  It is likely that as $N$ increases more constraints are needed. It would be interesting to study the importance and number of constraints on the distributions that are needed in the asymptotically large-$N$ limit. 

We note that our work builds on a previously-introduced free-then-freeze protocol involving ``breathing" particles for jammed packings~\cite{brito2018universality,brito2018theory}. As in our work, the learning degrees of freedom are the particle radii. However, in the case of breathing particles, each of the $N$ radii, or learning degrees of freedom, is subject to a constraint. In our case, we place the minimal number of constraints, far fewer than $N$, on moments of the learning-degree-of-freedom distributions. This allows us to reach far more stable local minima, making our free-then-freeze protocol more effective.

The introduction of transient learning degrees of freedom provides a unifying conceptual framework for the evolution of a variety of systems in complex energy or fitness landscapes. Appropriate training protocols can lead the disordered materials to occupy desired metastable states with useful properties 
~\cite{goodrich2015principle, hexner2018linking,pashine2019directed, hexner2020periodic,rocks2017designing,rocks2019limits,hexner2020periodic,stern2020supervised}. These ideas can be generalized to processing in experimental systems where the learning degrees of freedom can be ordinary translational ones. For example, with the right deposition protocol, vapor-deposited glasses have been created that are exceedingly stable~\cite{swallen2007organic, kearns2008hiking, ediger2017perspective}. In the context of transient learning degrees of freedom, the particles at a free surface have more degrees of freedom than those within the bulk. These variables can be considered learning degrees of freedom that freeze as more particles are deposited and the particles become buried in the bulk. 

As another example, one can interpret aging in supercooled liquids and glasses in terms of transient learning degrees of freedom.  
Upon aging, the relaxation times increase dramatically so that these pathways are inaccessible at short times. 
The relaxation itself produces a separation of time scales~\cite{ediger1996supercooled, berthier2004time}. 
The system, as it ages, limits the possibility of using the degrees of freedom (\textit{i.e.}, the relaxation pathways) so that they can also be considered to be transient degrees of freedom. This raises the possibility of using them as learning degrees of freedom to introduce desired properties.   

The concept of additional transient learning degrees of freedom is useful for thinking about many different protocols that have been used to create novel function in systems that exist in rugged landscapes.  While in some cases this leads only to a reinterpretation of what was already known, it serves the important purpose of allowing one to think about what kinds of new and different variables would be useful as a means to manipulate matter in new ways. A particularly interesting possibility to explore would be to introduce particle \textit{shapes} as learning degrees of freedom. Because shape couples strongly to the Maxwell's count at the jamming transition, it could therefore lead to a large increase in stability. The learning degrees of freedom studied in this paper can also be used to train for other functions as distinct from increased stability.  For example, one could consider training for a particular force network or for creating an allosteric interaction between distant particles. It is not clear which set of new variables, radii or stiffnesses, would be better at training for this function. 
\subsection{Acknowledgements}
We thank R Cameron Dennis for assistance with the constrained minimization algorithms.  This work is supported by the Simons Foundation for the collaboration
Cracking the Glass Problem via awards 454939 (EIC), 454945 (AJL), 348126 (VFH and SRN), 454947 (MLM) and Investigator Awards 327939 (AJL) and 446222(MLM), as well as by the US Department of Energy, Office of Science, Basic Energy Sciences, under Grant DE-SC0020972 (for studying implications of training in biological systems, SRN).  The comparison with experimental systems was partially supported by the University of Chicago Materials Research Science and Engineering Center funded by the National Science Foundation under award number DMR-2011854.

\bibliography{DOFandStability}
\clearpage
\newpage
\section*{Supplemental Information} \label{sec:Supplements}

\subsection*{Impact of Freezing Degrees of Freedom on Stability} 

Here, we show that freezing any subset of existing degrees of freedom in a system of particles can lead to an increase in the stability of the system. The two measuring tools we use are the distance to next instability, $\delta P$, and the density of states $\mathcal{D}(\omega)$ as discussed in Fig.~\ref{fig:panel}.
To measure $\delta P$, we first break the Hessian matrix into two terms, one that includes the rigidity matrix, $H_s = \mathbf{C}^{T} \mathbf{C}$, and one that includes the prestress forces, $H_p$. We then uniformly increase the prestress term using a dimensionless coefficient that includes the relative change in the pressure given by Eq.~(\ref{eq:hessian}):

\begin{equation}
   \label{eq:hessian}
   H = H_{s} + \frac{\delta P + P}{P} \ H_p
\end{equation}

\begin{figure}[h!]
\centering
\includegraphics[trim=0cm 0cm 0cm 0cm, width=8cm]{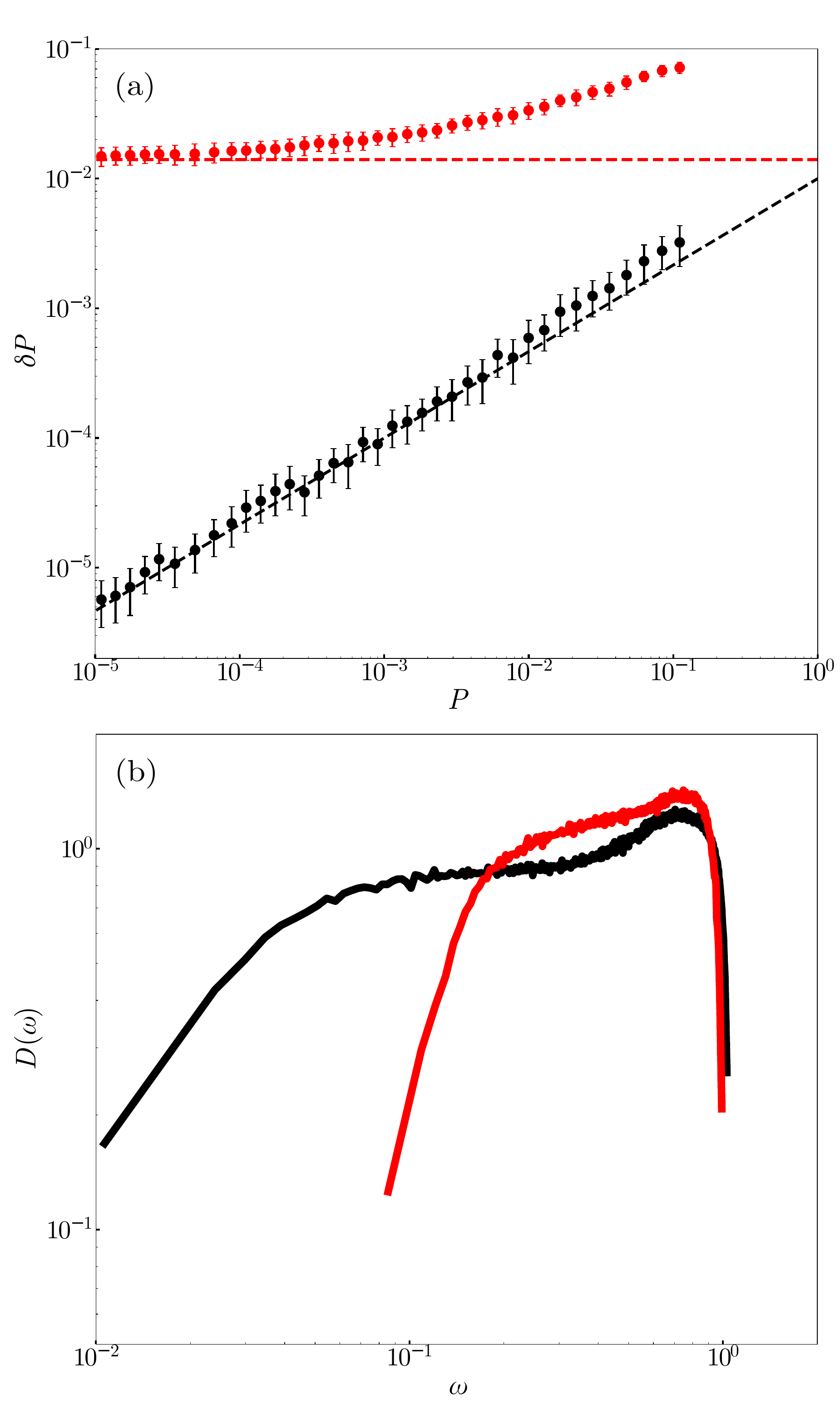}%
\caption{ a) The increase in pressure, $\delta P$, required to make a packing unstable before (black) and after freezing the x-components (red) of the positional degrees of freedom. The black dashed line show a power law $\delta P \propto P^{2/3}$. b) Density of states, $\mathcal{D}(\omega)$ versus $\omega$, for a packing before (black) and after freezing the x-components (red) of the positional degrees of freedom. 
All the data in (a) and (b) are ensemble averaged over $20$ mono disperse packings at pressure $p=10^{-4}$.}
\label{fig:freezingXDOF}
\end{figure}

Increasing $\delta P$ in Eq.~(\ref{eq:hessian}) is equivalent to increasing the contact forces uniformly. This pushes the system towards an instability without changing the geometrical configuration of the system. Since the prestress term in Eq.~(\ref{eq:hessian}) is negative definite, increasing its components will push the eigenvalues of the Hessian matrix, $H$, to zero. To measure the distance to an instability, we increase $\delta P$ in small steps and monitor the lowest non-zero eigenvalue of the Hessian. The $\delta P$ for which this eigenvalue goes to zero is data that is presented in Figs.~\ref{fig:panel}a,b, and~\ref{fig:freezingXDOF}a . 

Fig.~\ref{fig:freezingXDOF}a shows the change in the pressure, $\delta P$, required to push a system to a nearby instability when the motion of all the particles is confined in direction $x$ (red), in comparison to conventionally prepared packings (black) where the particles are free to move in all $d$ available directions. The result is very similar to Fig.~\ref{fig:panel}a, as freezing the $x$ degrees of freedom increases the distance to a nearby instability by a few orders of magnitude.

Fig.~\ref{fig:freezingXDOF}b shows the ensemble averaged density of states in $3D$ systems when the $x$ components of all particle positions are removed from the Hessian (red) compared to when all positional degrees of freedom are available (black). As can be seen from this plot, there is a clear shift to higher frequencies in the density of states which is similar to the shift in the lower tail of the density of states when radii were frozen at their equilibrium values in Fig.~\ref{fig:panel}c. Both plots in  Fig.~\ref{fig:freezingXDOF} show that freezing an existing subset of degrees of freedom can make the system more stable.

\subsection*{Impact of Radius and Stiffness Degrees of Freedom on Rigidity}
\label{differentDOF}

Here, we present a mathematical description for the difference between radius and stiffness degrees of freedom at the onset of rigidity. The main goal is to work out the type of degrees of freedom that can impact the Maxwell's count and shift the critical point. Since the Maxwell's count uses the rank-nullity theorem~\cite{alama2007rank+} on the rigidity matrix, we can start off by calculating this matrix which relates the changes in the constraints to changes in the available degrees of freedom~\cite{lubensky2015phonons, hagh2018rigidity}.
For instance, in a packing with $N$ particles and $N_c$ overlapping contacts, changes in the overlaps are directly related to changes in the particle positions through Eq.~(\ref{eqn:rigidity-overlap}):

\begin{equation}
\label{eqn:rigidity-overlap}
\mathbf{C} \mathbf{u} = \mathbf{\Delta}h
\end{equation}
where $\mathbf{u}$ is an $Nd \times 1$ dimensional vector with elements that represent the changes in the positions of particles $1$ through $N$, $\mathbf{C}$ is the rigidity matrix, and $\mathbf{\Delta}h$ is an $N_c \times 1$ dimensional vector where row $n$ represents the change in the overlap between the $n$th pair of particles when they move. This means that the rigidity matrix, $\mathbf{C}$, is an $N_c \times Nd$ dimensional matrix. 

Now, by adding the radius and stiffness degrees of freedom and a set of global constraints $\Phi_{\chi,m} = constant$, one can write a generalized version of Eq.~(\ref{eqn:rigidity-overlap}) in the form of $\mathbf{C} \mathbf{\Tilde{u}} = \mathbf{\Delta} h$ with $\mathbf{\Tilde{u}}$ is the vector of displacements in the $N(d+2) - \mu$ dimensional space of all degrees of freedom where $\mu$ is the number of globally applied constraints ($\mu = 7$ in this paper). The $\mathbf{\Delta}h$ vector is the differential change in the overlaps, which for a pair $ij$ is given by:

\begin{equation}
\label{eq:dh}
\begin{split}
dh_{ij} = 
\frac{\partial h_{ij}}{\partial x_i} dx_i + \frac{\partial h_{ij}}{\partial y_i} dy_i + \frac{\partial h_{ij}}{\partial z_i} dz_i +
\frac{\partial h_{ij}}{\partial R_i} dR_i + 
\frac{\partial h_{ij}}{\partial K_i} dK_i \\
+ 
\frac{\partial h_{ij}}{\partial x_j} dx_j + \frac{\partial h_{ij}}{\partial y_j} dy_j + \frac{\partial h_{ij}}{\partial z_j} dz_j +
\frac{\partial h_{ij}}{\partial R_j} dR_j + 
\frac{\partial h_{ij}}{\partial K_j} dK_j
\end{split}
\end{equation}

\begin{figure}[h!]
\centering
\includegraphics[width=6cm]{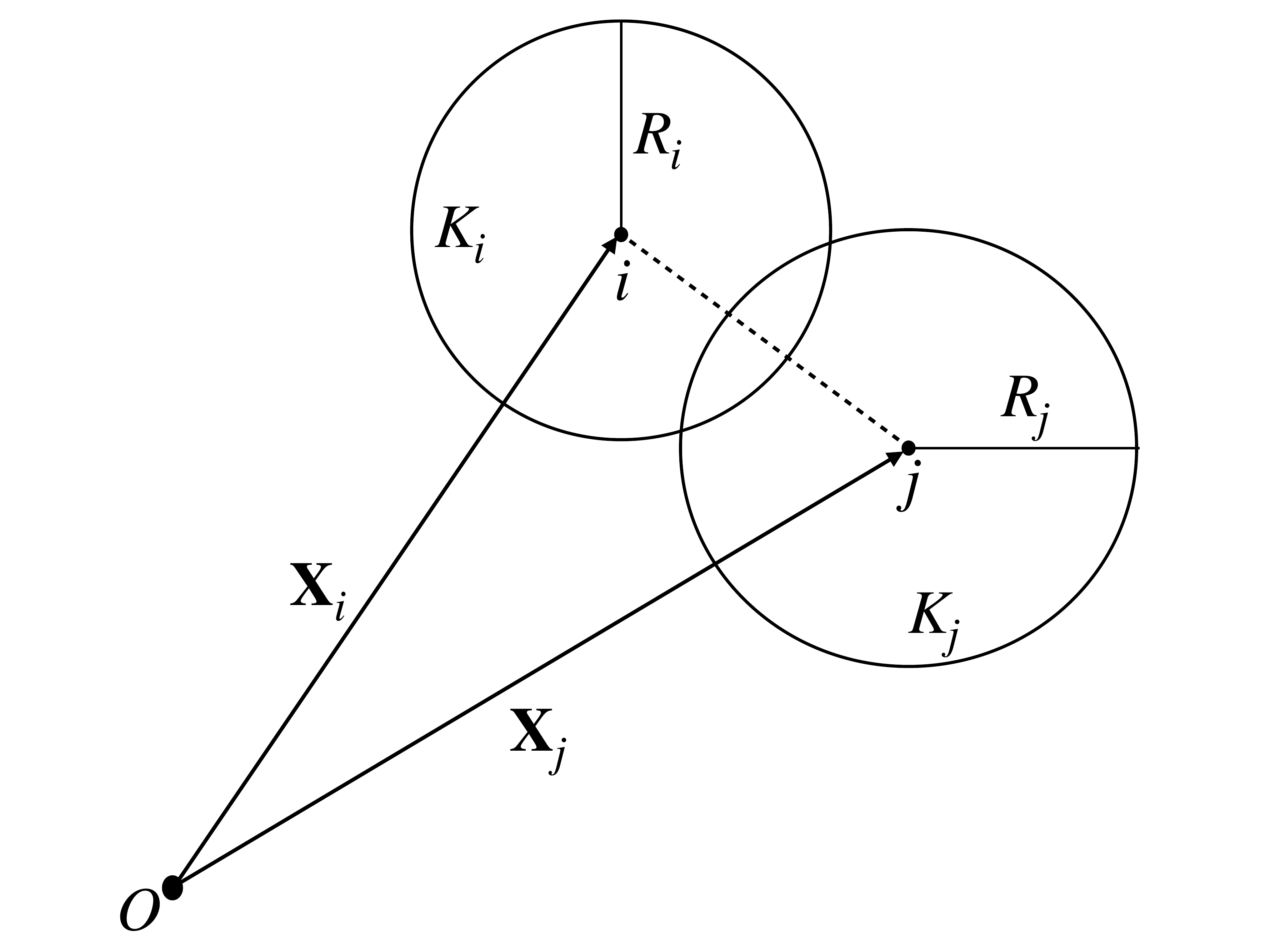}%
\caption{Two particles $i$ and $j$ with positions $\mathbf{X}_i$ and $\mathbf{X}_j$, radii $R_i$ and $R_j$, and stiffnesses $K_i$ and $K_j$ are shown. Changes in positions and radii of the particles can change their overlap.}
\label{fig:packingDis}
\end{figure}

The setup for such a pair is shown in Fig.~\ref{fig:packingDis}. Note that according to Eq.~(\ref{eq:energy}), the only variables that can change the overlaps and therefore the number of rows in $\mathbf{\Delta}h$ are the positions $\mathbf{\{ X_i \}}$ and radii $\{R_i\}$, since: 
\begin{equation}
\label{eqn:derivatives}
\begin{split}
\frac{\partial h_{ij}}{\partial x_i} & = \frac {-1}{R_i + R_j} \frac{x_i - x_j}{|\pmb{X}_i - \pmb{X}_j|}  \\
\frac{\partial h_{ij}}{\partial R_i}& = \frac{{|\pmb{X}_i - \pmb{X}_j|}}{(R_i + R_j)^2} = \beta_{ij} \\
\frac{\partial h_{ij}}{\partial K_i} & = 0 \\
\end{split}
\end{equation}

By writing the vector $\mathbf{\Tilde{u}}$ in the following form, 

\begin{align}
\label{eq:u}
\mathbf{\Tilde{u}} &= \begin{bmatrix}
    d x_1 \\
    d y_1 \\
    d z_1 \\
    \vdots \\
    d z_N \\
    d R_1 \\
    d R_2 \\
    d R_3 \\
    \vdots \\
    d R_{N- \mu} \\
    d K_1 \\
    d K_2 \\
    d K_3 \\
    \vdots \\
    d K_{N - \mu} \\
\end{bmatrix}
\end{align}
one can obtain the generalized rigidity matrix with positions, radii, and particle stiffnesses as degrees of freedom:
\begin{figure*}[htbp]
\begin{align*}
\label{eq:compatibility-generalized}
  \resizebox{.9\hsize}{!}{\bf{C} =
      \bordermatrix{
        & 1 & \hdots & i & \hdots & j & \hdots & N & \VR 
        & 1 & \hdots & i & \hdots & j & \hdots & N - \mu & \VR
        & 1 & \hdots & i & \hdots & j & \hdots & N - \mu & \cr
        \vdots & \vdots & \ddots & \vdots & \ddots & \vdots & \ddots & \vdots & \VR & \vdots & \ddots & \vdots & \ddots & \vdots & \ddots & \vdots & \VR & \vdots & \ddots & \vdots & \ddots & \vdots & \ddots & \vdots & \cr
      (i,j) & \bf{0} & \hdots & \bf{n}_{ij} & \hdots & \bf{n}_{ji} & \hdots & \bf{0} & \VR & 0 & \hdots & \beta_{ij} & \hdots & \beta_{ji} & \hdots & 0 & \VR & 0 & \hdots & 0 & \hdots & 0 & \hdots & 0 & \cr
      \vdots & \vdots & \ddots & \vdots & \ddots & \vdots & \ddots & \vdots & \VR & \vdots & \ddots & \vdots & \ddots & \vdots & \ddots & \vdots & \VR & \vdots & \ddots & \vdots & \ddots & \vdots & \ddots & \vdots & \cr
      }}
\end{align*}
\end{figure*}

where $\mathbf{n}_{ij} = \frac {-1}{R_i + R_j} \frac{\mathbf{X}_i - \mathbf{X}_j}{|\mathbf{X}_i - \mathbf{X}_j|} = -\mathbf{n}_{ji}$ and we assume that the number of global constraints on both the radius and stiffness degrees of freedom is equal to $\mu$. 

As can be seen from Eq.~(\ref{eq:compatibility-generalized}), adding the radii as degrees of freedom can change the rank of the rigidity matrix, $\text{rank}(\mathbf{C})$, but adding the particle stiffnesses as degrees of freedom does not change $\text{rank}(\mathbf{C})$ since it only adds $N - \mu$ extra columns of zeros to the matrix.

\subsection{Maxwell's count}
Maxwell's count is the result of rank-nullity theorem on the rigidity matrix: 

\begin{equation}
\label{eq:rank-nullity}
\begin{split}
\text{rank}(\mathbf{C}) + \text{nullity}(\mathbf{C}) & = \text{number of columns in } \mathbf{C} \\
\text{rank}(\mathbf{C^T}) + \text{nullity}(\mathbf{C^T}) & = \text{number of columns in } \mathbf{C^T} \\
\end{split}
\end{equation}
$\text{rank}(\mathbf{C}) = \text{rank}(\mathbf{C^T})$ is the number of independent rows/columns in $\mathbf{C}$ or $\mathbf{C^T}$. $\text{Nullity}(\mathbf{C})$ is the number of non-zero solutions to $\mathbf{C \Tilde{u} = 0}$. This is the number of trivial rigid motions that do not change the overlaps and therefore is equal to the total number of zero modes or floppy modes, $F$, in the system. $\text{Nullity}(\mathbf{C^T})$ on the other hand, is the number of non-zero solutions to $\mathbf{C^T f = 0}$, which is equal to the number of states of self-stress, $SSS$. This is because $\mathbf{C^T f}$ gives the vector of forces on the particles and therefore solutions to $\mathbf{C^T f = 0}$ represent all the possible contact forces that keep the system at equilibrium. On the right hand side of the first equation in~(\ref{eq:rank-nullity}), we have $\text{number of columns in } \mathbf{C} $ which is equal to the number of degrees of freedom $Nd + 2 (N - \mu)$. On the right hand side of the second equation in~(\ref{eq:rank-nullity}), we have $\text{number of columns in} \mathbf{C^T}$ which is equal to the number of contacts $N_c$. By subtracting the two rows in Eq.~(\ref{eq:rank-nullity}), one can write:

\begin{equation}
\label{eqn:Maxwell}
F = Nd + 2(N - \mu) - N_c + SSS
\end{equation}
which is the Maxwell's count for any system with position, radius, and stiffness degrees of freedom, $N_c$ contacts, and $\mu$ global constraints. The number of contacts, $N_c$, can also be written as $N_c = Z N/2$ where $Z$ is the average number of contacts per particle.
The isostatic point is defined as the critical point in which the number of degrees of freedom is balanced by the number of constraints and there are no states of self-stress, meaning that $F = d$ and $SSS = \text{nullity}(\mathbf{C^T}) = 0$. In other words:

\begin{equation}
\label{eqn:isostatic_z}
\begin{split}
\frac{Z_c N}{2} &=  \text{rank}(\mathbf{C^T}) \\
& = \text{rank}(\mathbf{C}) \\
& =  \text{number of dof} - \text{nullity}(\mathbf{C})\\
\end{split}
\end{equation}

\subsection{Maxwell's count when positions are the only degrees of freedom}
When the only degrees of freedom are positions, $\text{number of dof} = Nd$ and $\text{nullity}(\mathbf{C}) = d$. Therefore:

\begin{equation}
\label{eqn:isostatic_pos}
\frac{Z_c N}{2} = Nd - d
\end{equation}
which in the limit $N \rightarrow \infty$ gives $Z_c = 2d$.
\newline
\subsection{Maxwell's count when radii are added as new degrees of freedom}

Introducing radii as new degrees of freedom and $\mu$ global constraints adds $N - \mu$ new columns to the rigidity matrix and adds $1$ to the nullity of the rigidity matrix: 

\begin{equation}
\label{eqn:isostatic_rad}
\frac{Z_c N}{2} = N (d+1) - \mu - (d+1)
\end{equation}
which in the limit $N \rightarrow \infty$ gives $Z_c = 2(d+1)$.
This means that radii can move the critical point to a higher value as long as $\mu$ is negligible compared to $N$.

\subsection{Maxwell's count when stiffness of particles are added as new degrees of freedom}

Introducing stiffnesses as new degrees of freedom and $\mu$ global constraints, adds $N - \mu$ new columns to the rigidity matrix. But since all these columns are zero, it does not change the rank of the rigidity matrix. Based on the rank-nullity theorem, this means that stiffness degrees of freedom add $N - \mu$ new zero modes to the system and increase the $\text{nullity}(\mathbf{C}) $ by $N - \mu$:

\begin{equation}
\label{eqn:isostatic_stiff}
\begin{split}
\frac{Z_c N}{2} &= N (d+1) -\mu - (d+N -\mu) \\
& = Nd - d
\end{split}
\end{equation}
which in the limit $N \rightarrow \infty$ gives $Z_c = 2d$. This means that stiffness degrees of freedom do change the critical point simply because they do not appear in the definition of particle overlaps.

\end{document}